\documentclass[traditabstract]{aa}
\usepackage{epsfig,graphicx,natbib,txfonts,url,twoopt}
\usepackage[breaklinks=true]{hyperref} 
\usepackage[usenames]{color}

\def\@ #1\par{\par\mbox{}\\ \noindent{\small \tt @ #1}\\[1ex]}  

\bibpunct{(}{)}{;}{a}{}{,}    
\def\cite#1{\citealp{#1}}     
\newcommandtwoopt{\citeads}[3][][]{\href{http://adsabs.harvard.edu/abs/#3}%
                                        {\citealp[#1][#2]{#3}}}
\newcommandtwoopt{\citepads}[3][][]{\href{http://adsabs.harvard.edu/abs/#3}%
                                         {\citep[#1][#2]{#3}}}
\newcommandtwoopt{\citetads}[3][][]{\href{http://adsabs.harvard.edu/abs/#3}%
                                         {\citet[#1][#2]{#3}}}
\newcommandtwoopt{\citeyearads}%
  [3][][]{\href{http://adsabs.harvard.edu/abs/#3}%
  {\citeyear[#1][#2]{#3}}}

\newcount\longrefs
\def\aap{\ifnum\longrefs=1 {Astron.\ Astrophys.}\else 
                           {A\hbox{\rm \&}A}\fi}
\def\aapr{\ifnum\longrefs=1 {Astron.\ Astrophys.\ Rev.}\else 
                            {A\hbox{\rm \&}AR}\fi}
\def\aaps{\ifnum\longrefs=1 {Astron.\ Astrophys.\ Suppl.}\else 
                            {A\hbox{\rm \&}A Suppl.}\fi}
\def\aipcs{\ifnum\longrefs=1 {Am.\ Inst.\ Phys.\ Conf.\ Series}\else
                             {AIP Conf.\ Ser.}\fi}
\def\aj{\ifnum\longrefs=1 {Astron.\ J.}\else 
                          {AJ}\fi} 
\def\ao{\ifnum\longrefs=1 {Applied Optics}\else 
                           {Appl.\ Opt.}\fi} 
\def\aspcs{\ifnum\longrefs=1 {Astron.\ Soc.\ Pacific Conf.\ Series}\else 
                           {ASP Conf.\ Ser.}\fi} 
\def\apj{\ifnum\longrefs=1 {Astrophys.\ J.}\else 
                           {ApJ}\fi} 
\def\apjl{\ifnum\longrefs=1 {Astrophys.\ J. Lett.}\else 
                            {ApJ}\fi} 
\def\aplett{\ifnum\longrefs=1 {Astrophys.\ J. Lett.}\else 
                            {ApJ}\fi} 
\def\apjs{\ifnum\longrefs=1 {Astrophys.\ J. Suppl.}\else 
                            {ApJS}\fi}
\def\apss{\ifnum\longrefs=1 {Astrophys.\ and Space Science}\else 
                            {Astrophys.\ Space Sci.}\fi}
\def\araa{\ifnum\longrefs=1 {Ann.\ Rev.\ Astron.\ Astrophys.}\else 
                            {ARA\hbox{\rm \&}A}\fi}
\def\azh{\ifnum\longrefs=1 {Astronomicheskii Zhurnal}\else 
                            {Astron.\ Zhur.}\fi}
\def\baas{\ifnum\longrefs=1 {Bull.\ Am.\ Astron.\ Soc.}\else 
                            {BAAS}\fi}
\def\bain{\ifnum\longrefs=1 {Bull.\ Astronom.\ Institutes Netherlands}\else
                            {Bull.\ Astr.\ Inst.\ Neth.}\fi}
\def\gca{\ifnum\longrefs=1 {Geochim.\ Cosmochim.\ Acta}\else 
                           {Geochim.\ Cosmochim.\ Acta}\fi}
\def\grl{\ifnum\longrefs=1 {Geophys.\ Res.\ Lett.}\else 
                           {Geoph.\ Res.\ Lett.}\fi}
\def\iaucirc{\ifnum\longrefs=1 {IAU Circulars}\else 
                          {IAU Circ.}\fi}
\def\ip{\ifnum\longrefs=1 {in press}\else 
                          {in press}\fi}
\def\jgr{\ifnum\longrefs=1 {J.\ Geophys.\ Res.}\else 
                           {J.\ Geophys.\ Res.}\fi}  
\def\jrasc{\ifnum\longrefs=1 {J.\ Royal Astron.\ Soc.\ Canada}\else 
                           {JRAS Can.}\fi}  
\def\memsai{\ifnum\longrefs=1 {Mem.~Soc.~Astron.~Italiana}\else
                              {MemSAI}\fi}
\def\mnras{\ifnum\longrefs=1 {Mon.\ Not.\ Roy.\ Astron.\ Soc.}\else 
                             {MNRAS}\fi} 
\def\nat{\ifnum\longrefs=1 {Nature}\else 
                           {Nat}\fi}
\def\pasj{\ifnum\longrefs=1 {Pub.\ Astron.\ Soc.\ Japan}\else 
                            {PASJ}\fi} 
\def\pasp{\ifnum\longrefs=1 {Pub.\ Astron.\ Soc.\ Pacific}\else 
                            {PASP}\fi} 
\def\physscr{\ifnum\longrefs=1 {Physica Scripta}\else 
                            {Phys.\ Scrip.}\fi} 
\def\planss{\ifnum\longrefs=1 {Planetary \& Space Science}\else 
                            {Plan. \& Space Sci.}\fi} 
\def\procspie{\ifnum\longrefs=1 {Proc.\ SPIE}\else 
                            {Proc.\ SPIE}\fi} 
\def\qjras{\ifnum\longrefs=1 {Quarterly J.\ Royal Astron.\ Soc.}\else 
                            {QJRAS}\fi} 
\def\sa{\ifnum\longrefs=1 {Soviet Astron..}\else 
                               {Sov.\ Astron.}\fi}
\def\skytel{\ifnum\longrefs=1 {Sky \& Telescope}\else 
                            {Sky \& Tel.}\fi} 
\def\solphys{\ifnum\longrefs=1 {Solar Phys.}\else 
                               {Sol.\ Phys.}\fi}
\def\ssr{\ifnum\longrefs=1 {Space Science Rev.}\else 
                               {Space\ Sci.\ Rev.}\fi}
\def\zap{\ifnum\longrefs=1 {Zeitschr.\ f.\ Astrophysik}\else
                               {Z.\ Astrophys.}\fi}

\def\nl{,\ } 

\def\ITA{Institute of Theoretical Astrophysics\nl
         University of Oslo\nl
         P.O. Box 1029, Blindern\nl N--0315 Oslo\nl Norway}

\def\LMSAL{Lockheed-Martin Solar and Astrophysics Laboratory\nl
           3251 Hanover Street\nl Palo Alto, CA~94304\nl USA}

\def\SIU{Sterrekundig Instituut\nl Utrecht University\nl Postbus 80\,000\nl
         NL--3508~TA~Utrecht\nl The~Netherlands}

\long\def\startignore #1\stopignore{}   

\def\rmit#1{{\it #1}}              
\def\etal{\rmit{et al.}}           
           
\def\ie{\rmit{i.e.,}}              
\def\eg{\rmit{e.g.,}}              
\def\cf{cf.}                       

\def\specchar#1{\uppercase{#1}}    

\def\CI{\mbox{C\,\specchar{i}}}

\def\CaII{\mbox{Ca\,\specchar{ii}}}

\def\FeI{\mbox{Fe\,\specchar{i}}} 
\def\FeII{\mbox{Fe\,\specchar{ii}}}

\def\HI{\mbox{H\,\specchar{i}}} 
 
\def\Hmin{\hbox{\rmH$^{^{_-}}\!$}}      

\def\MgI{\mbox{Mg\,\specchar{i}}} 
\def\MgII{\mbox{Mg\,\specchar{ii}}} 
\def\MgIII{\mbox{Mg\,\specchar{iii}}} 
\def\MgIV{\mbox{Mg\,\specchar{iv}}}

\def\SiI{\mbox{Si\,\specchar{i}}} 
\def\SiII{\mbox{Si\,\specchar{ii}}}

\def\Halpha{\mbox{H\hspace{0.1ex}$\alpha$}} 

\def\Hepsilon{\mbox{H\hspace{0.2ex}$\epsilon$}}
\def\Lyalpha{\mbox{Ly$\hspace{0.2ex}\alpha$}}

\def\CaIIH{\mbox{Ca\,\specchar{ii}\,\,H}}

\def\HK{\mbox{H\,\&\,K}}
\def\Kthree{\mbox{K$_3$}}      

\def\KtwoV{\mbox{K$_{2V}$}}

\def\HtwoV{\mbox{H$_{2V}$}}

\def\hk{\mbox{h\,\&\,k}}

\def\level #1 #2#3#4{$#1 \; ^{#2} \mbox{#3} ^{#4}$}

\def\rmd{{\rm d}}  
\def\rme{{\rm e}}

 \def\rmH{{\rm H}}

\def\kms{\hbox{km$\;$s$^{-1}$}}

\def\is{\!=\!}                             
\def\lambdop{\hbox{$\bf \Lambda$}}         

\def\={\hbox{$\!=\!$}}                     

\def\rmit#1{#1}               
\def\specchar#1{{\sc #1}}     

\def\figspath{.}

\def\helium{He\,II\,304\,\AA}
\def\CaIR{Ca\,II\,8542\,\AA}
\def\fontenla{\citetads{2009ApJ...707..482F}}

\def\fonmod{FCHHT-B}

\begin{document} 

\title{\mbox{}\vspace{3mm}\newline 
Chromospheric backradiation in ultraviolet continua and H$\alpha$ 
     }
\subtitle{}
\titlerunning{Chromospheric backradiation}

\author{R.J. Rutten\inst{1,2,3}\thanks{Address: 
Lingezicht Astrophysics, 't Oosteneind 9, 4158CA Deil, The Netherlands}
        \and
        H. Uitenbroek\inst{4}}

\authorrunning{R.J. Rutten \etal}

\institute{\SIU \and \ITA \and \LMSAL \and 
National Solar Observatory\,/\,Sacramento Peak\thanks{Operated by the
    Association of Universities for Research in Astronomy,
    Inc. (AURA), for the National Science Foundation},
    P.O.~Box 62, Sunspot, NM 88349, USA}

\date{Received 27 November 2011 / Accepted 29 February 2012}
\offprints{R.J. Rutten\\  
\email{R.J.Rutten@uu.nl}}

\abstract{A recent paper states that ultraviolet backradiation from
  the solar transition region and upper chromosphere strongly affects
  the degree of ionization of minority stages at the top of the
  photosphere, \ie\ in the temperature minimum of the one-dimensional
  static model atmospheres presented in that paper.
  We show that this claim is incompatible with observations and we
  demonstrate that the pertinent ionization balances are instead
  dominated by outward photospheric radiation, as in
  older static models.  We then analyze the formation of \Halpha\
  in the above model and show that it has significant
  backradiation across the opacity gap by which \Halpha\ differs
  from other strong scatttering lines.
\keywords{Sun: photosphere -- Sun: chromosphere --
          Sun: UV radiation}}

\maketitle

\section{Introduction}                             \label{sec:introduction}
Static one-dimensional (1D) modeling of the solar atmosphere assumes
hydrostatic equilibrium in plane-parallel-layer geometry to deliver
temperature and density stratifications along vertical columns that
may serve as reference description to compute spectral continua and
lines similar to actual solar radiation.  Early versions of such
``standard'' models were compiled by
\citetads{1959HDP....52...80D} 
and \citeauthor{1964BAN....17..442H} 
(\citeyearads{1964BAN....17..442H}, 
\citeyearads{1964SAOSR.167..240H}). 
Holweger's (\citeyearads{1967ZA.....65..365H}) 
LTE best-fit to optical iron lines became the first choice in
classical abundance determination, in particular as the HOLMUL update
by \citetads{1974SoPh...39...19H}. 
This model describes only the photosphere and is close to theoretical
photosphere models based on the additional assumption of energy
conservation in the form of radiative equilibrium plus mixing-length
convection in the deepest layers (\eg\
\citeauthor{1974SoPh...34...17K} 
\citeyearads{1974SoPh...34...17K}, 
\citeyearads{1994KurCD..19.....K}; 
\citeauthor{1975A&A....42..407G} 
\citeyearads{1975A&A....42..407G}, 
\citeyearads{2008A&A...486..951G}). 
Semi-empirical best-fit modeling of solar continua including
ultraviolet wavelengths sampling the chromosphere continued with the
Bilderberg Continuum Atmosphere
(\citeads{1968SoPh....3....5G}), 
the Harvard-Smithsonian Reference Atmosphere
(\citeads{1971SoPh...18..347G}), 
the VAL model grids of \citeauthor{1973ApJ...184..605V}
(\citeyearads{1973ApJ...184..605V}, 
\citeyearads{1976ApJS...30....1V}, 
\citeyearads{1981ApJS...45..635V}) 
fitting disk-center Lyman brightness bins,
the update by \citetads{1986ApJ...306..284M}, 
the more recent update concentrating on disk-center ultraviolet
spectra by \citetads{2008ApJS..175..229A}, 
and the sequence of similar refinements by
\citeauthor{1990ApJ...355..700F} 
(\citeyearads{1990ApJ...355..700F}, 
\citeyearads{1991ApJ...377..712F}, 
\citeyearads{1993ApJ...406..319F}, 
\citeyearads{2002ApJ...572..636F}, 
\citeyearads{2006ApJ...639..441F}, 
\citeyearads{2007ApJ...667.1243F}, 
\citeyearads{2009ApJ...707..482F}). 

This paper is triggered by the conclusion of \fontenla\ that {\em ``in
  the solar chromosphere, the FUV and EUV observed emissions in
  continuum and lines produced in the upper chromosphere and
  transition region irradiate the low chromosphere and have
  significant effects on the ionization.''}.  In particular, they
claim that such backradiation strongly affects the density of minority
stages of ionization in and near the temperature minimum of their
quiet-Sun model.  We first show that this claim is incompatible with
basic observations, and then use illustrative computations to
demonstrate that, instead, photospheric irradiation from below is the
dominant agent in ultraviolet continuum formation in the
quiet-Sun model of \fontenla, as was the case for the older standard
models.  

We continue by demonstrating that significant chromospheric
backradiation occurs in \Halpha\ in such static models through
two-level scattering (Sect.~\ref{sec:Halpha}), but argue that the
assumption of columnar hydrostatic equilibrium is untenable for the
chromosphere and transition region (Sect.~\ref{sec:discussion}).

\def\figlabel{fig:DOT-SDO}  
\begin{figure*}      
  \centering
  \includegraphics[width=18cm]{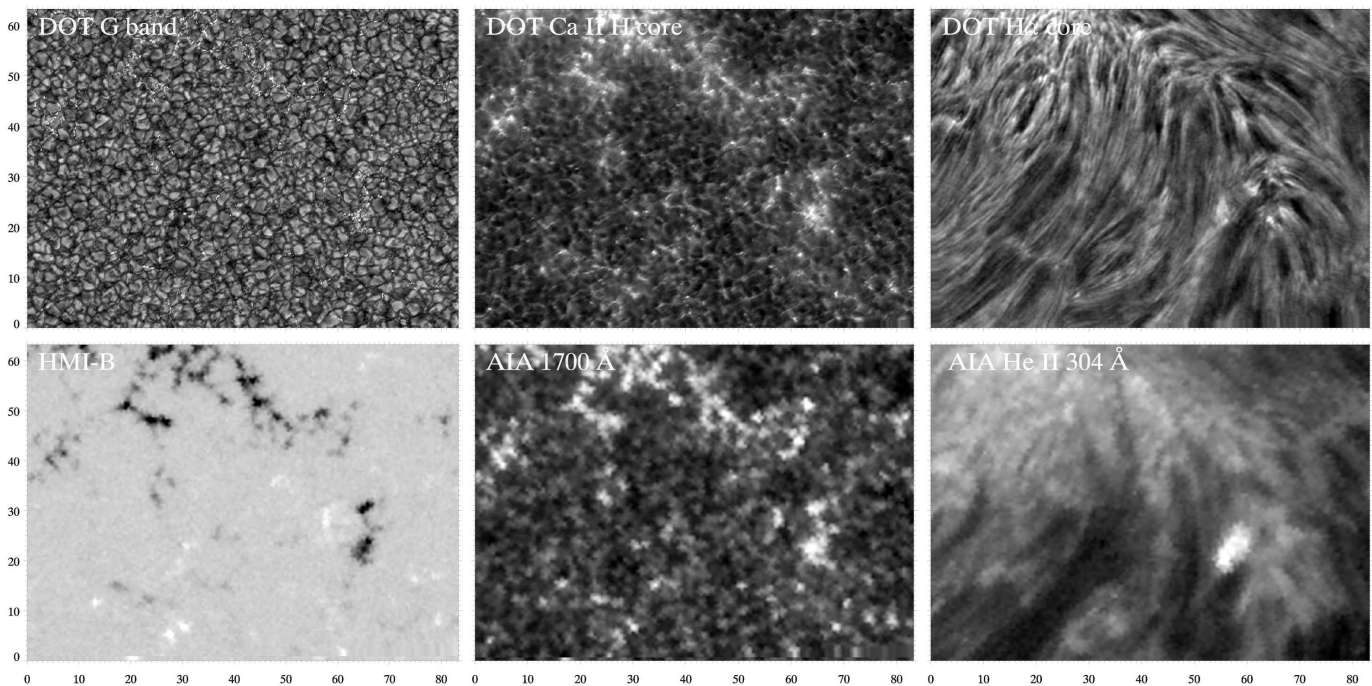}  
  \caption[]{\label{\figlabel} \footnotesize %
    Co-spatial simultaneous images from the Dutch Open Telescope (DOT)
    at \url{ftp://dotdb.strw.leidenuniv.nl} and from the
    Solar Dynamics Observatory (SDO) at
    \url{http://lmsal.com/get_aia_data}.  First row: DOT images
    sampling the G band, the core of \CaIIH\ with 1.4~\AA\
      passband, the center of \Halpha\ with 0.25~\AA\ passband.
    Second row: SDO/HMI magnetogram, SDO/AIA 1700\,\AA\ image, SDO/AIA
    \helium\ image.  They were all taken on September 27, 2010 within
    a few seconds around 13:54:30\,UT, a moment of excellent seeing at
    the DOT (Fried parameter $r_0 \!  \approx \! 14$\,cm).  The field
    of view (scales in arcsec, terrestrial North up) is near disk
    center, South of the largest sunspot of AR11109.  It contains a
    ridge of plage in its upper half, quieter areas in its lower half.
    The greyscales have been optimized by clipping (all images) and
    logarithmic scaling (AIA images).  We invite the reader to zoom in
    with a pdf viewer for better appreciation of the detail in the DOT
    images.  DOT observer: E.~Romashets. DOT speckle reconstruction:
    A.~Sukhorukov.  Alignment routines: P.~S\"utterlin.}
\end{figure*}

\section{Observation}                  \label{sec:observation}
In this section we briefly review the observational character of
the chromosphere, as counterpart to the standard-model definition as
the domain between the model's temperature minimum and steep
temperature rise towards the corona.  Figure~\ref{fig:DOT-SDO} gives
an overview of the appearance of the solar atmosphere in an area
containing plage and network.  We use it to demonstrate that
the actual chromosphere may have enough opacity to supply
significant backradiation in \Halpha, but not in the 1700\,\AA\
continuum.

The G-band image and the magnetogram in the first column of
Fig.~\ref{fig:DOT-SDO} are photospheric.  The first image displays the
granulation and the spatial distribution of small kilogauss magnetic
concentrations qualitatively and without sign through their proxy as
intergranular bright points (best seen when zooming in).  The second
image quantifies the actual surface distribution of the
line-of-sight component in such strong-field concentrations, but
at lower angular resolution.  The correspondence is excellent.

The other four images allegedly sample the chromosphere (1700\,\AA,
\CaIIH, \Halpha) and the transition region (\helium).  However, only
the \Halpha\ image truly shows the chromosphere.  It consists of long
slender fibrils that cover and obscure the photosphere
everywhere in the form of overlying fibrilar canopies.  This is
  always the case near active regions, plage, and active network; only
  the very quietest internetwork areas are not covered by fibrils in
  \Halpha\ (\eg\ Plate 3.1 of
  \citeads{1974soch.book.....B}; 
  \citeads{2007ApJ...660L.169R}; 
  \citeads{2008SoPh..251..533R}). 
  These ubiquitous \Halpha\ fibril canopies constitute the ring of
pink emission just off the limb around the Sun that
\citetads{1868RSPS...17..131L} 
named ``chromosphere''.  They are also likely to irradiate the
  low-opacity domain underneath with \Halpha\ backradiation.  Their geometry
  remains unknown; the fibrilar appearance in images as this one may
  represent long slender cylindrical fluxtubes 
  (\eg\ \citeads{1994A&A...282..939H}), 
  density variations that cause corrugations of the $\tau \is 1$
  Eddington-Barbier surface in an extended atmosphere
  (\citeads{1972SoPh...22..344S}), 
  or warps in sheets as proposed by 
  \citetads{2011ApJ...730L...4J}. 

The 1700\,\AA\ continuum shows no fibril canopies covering
internetwork areas but the underlying domain of cool
gas that is every few minutes ridden through by successive hot shocks
producing \CaII\ \HtwoV\ and \KtwoV\ cell grains
(\citeads{1991SoPh..134...15R}; 
 \citeads{1997ApJ...481..500C}; 
\cf\ \citeads{2011arXiv1110.6606R}). 
It is called ``clapotisphere'' here following
\citetads{1995ESASP.376a.151R} 
and is badly represented by the hydrostatic standard models
(\citeads{1995ApJ...440L..29C}). 
The bright points marking magnetic concentrations and making up the
network and plage are roughly co-spatial with the deeper G-band bright
points.  Their large brightness contrast represents a mixture of
scattered hot-wall radiation from below and current heating
(\citeads{2010MmSAI..81..582C}). 
Numerical simulations suggest that, due to the latter heating,
magnetic concentrations do contain temperature stratifications
resembling the standard models (\eg\ the FAL-C model of
\citeads{1993ApJ...406..319F}), 
but with the temperature minimum shifted down to height $h \!\approx\!
150$\,km, still within the 
photosphere, due to fluxtube evacuation (first quartet in Fig.~9 of
\citeads{2010ApJ...709.1362L}). 
Ubiquitous shocks occur also in and near these concentrations, but
also already deeper than in the shocked internetwork clapotisphere
(\citeads{1998ApJ...495..468S}; 
\citeads{2011A&A...531A..17R}). 
Slow hydrogen ionization/recombination balancing in the cool
aftermath of the shocks causes large NLTE overpopulation of the lower
level of \Halpha\
(\citeads{2007A&A...473..625L}). 

The \CaIIH\ image is much like the 1700\,\AA\ image, apart from
difference in angular resolution.  In the internetwork it shows a
bright mesh pattern that is dominated by shock interference and
changes rapidly.  There are no chromospheric fibrils.  One should
rather expect that fibrils that are opaque in \Halpha\ are
also opaque at the center of \CaIIH, because the thicker parts
of \Halpha\ fibrils are seen also in \CaIR\
(\citeads{2009A&A...503..577C}), 
and \CaIIH\ is bound to exceed \CaIR\ in extinction.  Indeed, in
the VAL3-C model \CaII\ \Kthree\ forms higher than the core
of \Halpha\ (Fig.~1 of \citeads{1981ApJS...45..635V}). 
However, filtergrams such as this one are made with wide passbands
(1.4\,\AA\ for the DOT) so that the internetwork scene is dominated by
shock brightness in the inner line wings that swamps the signature of
very dark fibrils at line center
(\citeads{2009A&A...500.1239R}). 
Compared to the 1700\,\AA\ image, there is an extra contribution in
the form of a diffuse bright haze around network and plage that is
likely the unresolved appearance of the highly dynamic features
seen as \CaIIH\ straws near the limb
(\citeads{2006ASPC..354..276R}), 
\CaIIH\ spicules-II outside the limb
(\citeads{2007PASJ...59S.655D}), 
and RBEs (rapid blue excursions) on the disk in the blue wing of
\Halpha\ (\citeads{2009ApJ...705..272R}). 

The \helium\ image shows fibrilar patterns that are not identical to
but also not unlike the \Halpha\ canopies.  It also shows similar
bright grains near fibril feet as \Halpha\ does.  These are not
cospatial with the bright grains in the G-band and \CaIIH\ images.
The overall similarity suggests that hot transition-region gas shares
the field-guided topography of the \Halpha\ chromosphere, outlining
similar magnetic connectivity patterns.
Such patterns seem also to be mapped by \Lyalpha\ fibrils in VAULT
images (\citeads{2010SoPh..261...53V}). 
Any fibril that is opaque in \Halpha\ must be very thick in
\Lyalpha\ but, comparably to \CaIIH, the cell-covering fibrils are very
dark in \Lyalpha\, whereas other structures, likely Doppler-shifted out of
line-center obscuration, contribute larger brightness in the
profile-summed VAULT images
(\citeads{2009A&A...499..917K}). 

The conclusion from this section is that significant
chromospheric backradiation may be expected in \Halpha\ and in
\helium\ which show opaque fibrilar canopies that constitute the
actual chromosphere.  \CaIIH\ should show them too at line
  center, but the DOT image and all other \HK\ filtergrams have
too wide a bandpass.
In the 1700\,\AA\ image the absence of any fibrilar signature, bright
or dark or as erasure of the internetwork pattern seen in \CaIIH,
implies that they are transparent at this wavelength, even for such a
not-so-quiet area.  Since the backradiation by an optically thin
fibril scales with the fibril opacity, it is unlikely that deeper
layers are affected by chromospheric backradiation in such ultraviolet
continua.

\section{Formulation}                  \label{sec:formulation}

\fontenla\ based their claim of important chromospheric backradiation
on their Fig.~4 containing graphs of population departure coefficients
$b$ for selected levels of \SiI, \SiII, \MgI, MgII, \FeI, and \FeII.
Population departure coefficients measure the ratio of the actual
population density (particles\,cm$^{-3})$ to the population computed
assuming LTE.  There are two formats differing in normalization.  The
first, called ``Menzel'' here, follows the original use by
\citetads{1937ApJ....85...88M} 
who normalized hydrogen populations to the free proton density.  The
generalization is to use a partial Saha-Boltzmann evaluation to obtain
normalization to the next ion stage (Eq.~(6.3) of
\citeads{1968slf..book.....J}; 
Eqs.~(13)--(15) of \citeads{1981ApJS...45..635V}). 
The other format, called ``Zwaan'' here, follows
\citetads{1972SoPh...23..265W} 
and uses the elemental abundance as normalization.  The two
conventions are (\citeads{2003rtsa.book.....R}): 
\begin{equation}
  b_i^{\rm Zwaan} \equiv n_i / n_i^{\rm LTE} 
  \mbox{\rm ~~~~ and ~~~~}
  b_i^{\rm Menzel} \equiv \frac{n_i/n_i^{\rm LTE}}
                               {n_C/n_C^{\rm LTE}},
          \label{eq:bdef}
\end{equation}
where $n_i$ is the population of level $i$, $n_C$ is the total
population of the next ion, and the superscript LTE implies a complete
solution of the Boltzmann-Saha LTE partitioning equations for the
given level and all stages of the element.  Sometimes the ion ground
state $n_c$ is used instead of $n_C$ assuming $n_c \approx n_C$.  When
the atom or ion containing level $i$ is predominantly ionized, so that
$n_C \approx n_C^{\rm LTE}$ because most particles of the species are
in the next-ion stage, then
\begin{equation}
     b_i^{\rm Menzel} \approx n_i/n_i^{\rm LTE} \approx b_i^{\rm Zwaan},
    \label{eq:ionizedb}
\end{equation}
but when most of the element sits in level $i$ itself then
\begin{equation}
     b_i^{\rm Menzel} \approx n_C^{\rm LTE}/n_C \approx 1/b_C^{\rm Zwaan}.
    \label{eq:neutralb}
\end{equation} 
The continuum has $b_C^{\rm Menzel} \equiv 1$ and the ratio $b_i/b_C
\approx b_i/b_c$ is the same in both definitions.  

Equations~(\ref{eq:ionizedb}) and (\ref{eq:neutralb}) were given as
Eqs.~(17) and (18) by \citetads{1981ApJS...45..635V}, 
who used the Menzel definition.  \fontenla\ did so too (private
communication from J. Fontenla).  We show results for both definitions
here because the Zwaan version gives a more intuitive display of
actual population departures for majority stages.  For example, the
Menzel-coefficient dips for $n=1$ of \HI\ and \CI\ at the temperature
minimum in Figs.~30 and 33 of \citetads{1981ApJS...45..635V} 
and Figs.~13 and 14 of
\citetads{2007ApJ...667.1243F} 
do not imply that the populations of these levels are out of LTE.

\def\figlabel{fig:models}
\begin{figure}      
  \centering
  \includegraphics[width=75mm]{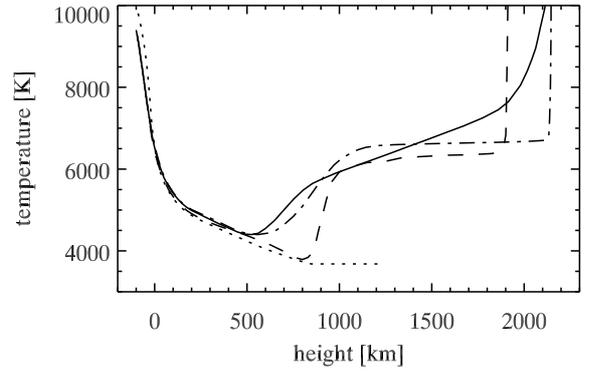} 
  \caption[]{\label{\figlabel} \footnotesize %
    Temperature stratification in some standard models.  {\em
      Solid\/}: FAL-C of
    \citetads{1993ApJ...406..319F}. 
    {\em Dot-dashed\/}: its AL-C7 update by
    \citetads{2008ApJS..175..229A}. 
    {\em Dashed\/}: \fonmod\ of \fontenla.  {\em Dotted\/}:
    radiative-equilibrium model from Kurucz
    (\citeyearads{1979ApJS...40....1K}, 
    \citeyearads{1992RMxAA..23..181K}, 
    \citeyearads{1992RMxAA..23..187K}), 
    extended to larger height assuming constant temperature. %
  }\end{figure}

\section{Demonstration}               \label{sec:demonstration}
We select magnesium for plotting departure coefficients as in Fig.~4
of \fontenla.  Magnesium is an important electron donor in the upper
photosphere where \MgI\ has larger Rydberg populations than any other
element (Fig.~15 of
\citeads{1992A&A...253..567C}). 
It has important bound-free edges, at 1621.5\,\AA\ (vacuum) from the
ground state (\level 3s^2 1Se{}) and especially at 2512.4\,\AA\ (air)
from the first excited level (\level 3s3p 3P{}).  \MgI\ and \MgII\
spectrum formation are characteristic for other abundant metals with
low ionization energy (silicon, iron, aluminum).  Together, these
supply the electrons for the photospheric \Hmin\ opacity in the
visible and infrared and they dominate the solar continuous opacity
from the near-ultraviolet to \Lyalpha\ (Fig.~36 of
\citeads{1981ApJS...45..635V}). 

Figure~\ref{fig:models} displays a selection of standard models.
FAL-C of \citetads{1993ApJ...406..319F} 
was an update of VAL3-C of \citetads{1981ApJS...45..635V} 
with a less steep upper photosphere from the inclusion of the
ultraviolet line haze and a different transition region from the
inclusion of ambipolar diffusion.  A more recent update is the AL-C7
model of
\citetads{2008ApJS..175..229A}, 
constructed in particular to reproduce the ultraviolet spectrum
atlases of \citetads{1993ApJS...87..443B} 
and \citetads{2001A&A...375..591C}. 
These spectra are also reproduced by model \fonmod\ of \fontenla,
which has a higher-located temperature minimum to accommodate dark
infrared CO lines.

The dotted curve in Fig.~\ref{fig:models} is a radiative-equilibrium
model from Kurucz (\citeyearads{1979ApJS...40....1K}, 
\citeyearads{1992RMxAA..23..181K}, 
\citeyearads{1992RMxAA..23..187K}), 
that we extended in \citetads{2009A&A...503..577C} 
to larger height assuming constant temperature.  It is very similar to
the \fonmod\ model up to $h\ \is 800$~km, but has no chromosphere or
transition region and so serves as comparison for the case of no
backradiation whatsoever.

We have used the 1D version of the RH code (\eg\
\citeads{2001ApJ...557..389U}) 
in a setup with similar content as the PANDORA setup of
\citetads{2008ApJS..175..229A}, 
although the solving method differs much.  RH uses a multi-level
accelerated lambda iteration method following
\citeauthor{1991A&A...245..171R}
(\citeyearads{1991A&A...245..171R}, 
\citeyearads{1992A&A...262..209R}) 
-- hence the code's name.  We used the 66-level model atom for \MgI\
of \citetads{1992A&A...253..567C} 
and a 12-level model atom for \MgII\ from
\citetads{1997SoPh..172..109U}. 
\MgIII\ is represented by its ground state only.  Other elements of
which the more important transitions are explicitly evaluated in NLTE
are H, Si, Al, and Fe.  Partial redistribution is accounted for in
\MgII\ \hk\ and \Lyalpha\ and $\beta$.  The violet and ultraviolet
line haze (\eg\ \citeads{1980A&A....90..239G}) 
causing a quasi-continuum is accounted for by sampling the 236\,000
lines between $\lambda \is 1000$ and 4000~\AA\ in the extensive list
of Kurucz\footnote{\url{http://kurucz.harvard.edu/linelists.html}}
every 20\,m\AA, using a two-level coherent scattering approximation
per wavelength in which the collisional transition probability of each
sampled line is estimated from the radiative one with the
approximation of
\citetads{1962ApJ...136..906V}, 
rather than applying a common scattering recipe for all lines as done
by \citetads{2008ApJS..175..229A}. 
However, a test with the much simpler line-haze recipe of
\citetads{1992A&A...265..237B} 
showed only little differences with the results shown here.

\def\figlabel{fig:fractions}
\begin{figure}      
  \centering
  \includegraphics[width=88mm]{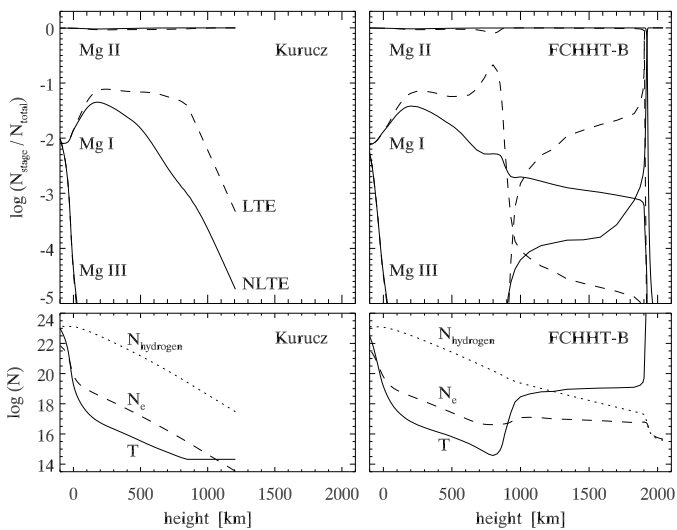}  
  \caption[]{\label{\figlabel} \footnotesize %
    {\em Upper panels\/}: magnesium ionization fractions $N_{\rm
      stage} / N_{\rm total}$ in the Kurucz model ({\em left\/}) and
    the \fonmod\ model ({\em right\/}), with $N_{\rm stage}$ the summed
    populations per ionization stage, $N_{\rm total}$ the total of all
    stages.  {\em Solid\/}: NLTE.  {\em Dashed\/}: LTE.  {\em Lower
      panels\/}: gradient comparisons.  The $y$-axis units specify the
    logarithmic electron density $N_\rme$ and total hydrogen density
    $N_\rmH$ in particles\,m$^{-3}$.  The temperature curves show
    $20\,\log(T)-57$ to provide comparable gradients. %
  }\end{figure}

Figure~\ref{fig:fractions} shows the resulting NLTE and LTE ionization
fractions for the Kurucz and \fonmod\ models.  In both magnesium is
virtually once-ionized throughout the atmosphere (below the transition
region at $h \is 1900$\,km for \fonmod).  Let us first interpret the
simple LTE curves in the upper Kurucz panel, which reflect the
competing effects in the Saha equation of the outward decreases of the
temperature $T$ and electron density $N_\rme$ shown in the lower
Kurucz panel.  Both decreases are steepest in the deepest layers,
where $N_\rme$ drops from $10^{-1}$ to $10^{-4}$ of the hydrogen
density $N_\rmH$ because hydrogen becomes neutral.  The $N_\rme /
N_{\rm H} \!\approx\!  10^{-4}$ limit represents the combined
abundance of the electron donor elements (Fe, Mg, Si, Al) and controls
the \Hmin\ opacity.  The steep $T$ drop wins from the steep
$N_\rme$ drop so that the \MgI\ fraction increases and the \MgIII\
fraction decreases with height.  Around $h \is 500$\,km the $N_\rme$
drop wins from the $T$ drop so that the LTE \MgI\ fraction drops
slightly.  The outer exponential decay of this fraction follows the
decreasing gas density (with constant $N_\rme / N_{\rm H} \!\approx\!
10^{-4}$) at constant temperature.  The NLTE \MgI\ curve also shows
near-exponential decay around $h \is 500$\,km, implying that the
nonthermal contribution, from radiative instead of collisional
ionization and recombination, is about constant with height.

The \MgI\ and \MgIII\ curves in the \fonmod\ panel are closely the same
as in the Kurucz panel up to the \fonmod\ temperature minimum at
$h\is800$\,km, just as the models are.  Thus, up to this height magnesium
ionization does not depend much on higher layers.

\def\figlabel{fig:bground}
\begin{figure}      
  \centering
  \includegraphics[width=88mm]{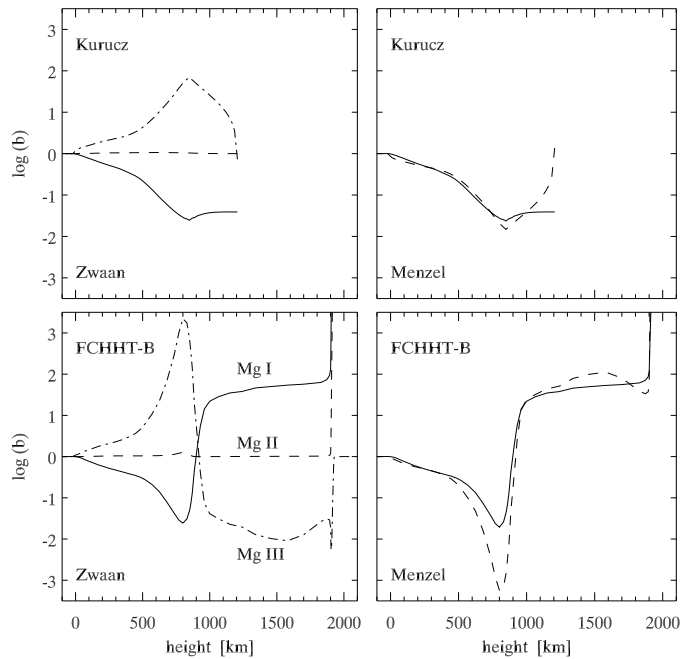}  
  \caption[]{\label{\figlabel} \footnotesize %
    Population departures of magnesium ground states for the
    Kurucz model ({\em upper panels\/}) and the \fonmod\ model ({\em
      lower panels\/}).  {\em Left\/}: Zwaan definition.  {\em
      Right\/}: Menzel definition.  {\em Solid\/}: \MgI\ ground state.
    {\em Dashed\/}: \MgII\ ground state.  {\em Dot-dashed\/}: \MgIII\
    ground state. %
  }\end{figure}

Figure~\ref{fig:bground} shows the population departures $b$ of the
magnesium ground states for the Kurucz and \fonmod\ models.
Both $b$ definitions are used (left and right columns).  \MgIII\
cannot be shown in the Menzel version because we do not include \MgIV.
Since most of the population per ionization stage resides in the
ground state, these curves reflect the divergences between the LTE and
NLTE ionization fractions in Fig.~\ref{fig:fractions}.

The \fonmod/Menzel curves in the last panel resemble the
corresponding curves in Fig.~4 of \fontenla, which for \SiII\ and
\FeII\ dip yet deeper than for \MgII.  At first sight the two deep
dips in this panel would suggest large depletion of both \MgI\ and
\MgII\ at the temperature minimum.  However, for \MgII\ this dip is
the reverse of the \MgIII\ peak in the Zwaan panel following
Eq.~(\ref{eq:neutralb}).  In the Zwaan panels the \MgII\ curves remain
near unity because even large overionization of \MgI\ and
overpopulation of \MgIII\ do not much affect the \MgII\
population, being dominant anyhow (Fig.~\ref{fig:fractions}).

Near and above the temperature minimum, the \MgI\ and \MgIII\ curves
in the first panel of Fig.~\ref{fig:bground} mimic the temperature
structure of the model atmosphere, for \MgIII\ reversely.  This
implies that the actual degree of ionization does not sense the
temperature but is constant with height, so that the departure
behavior is set by the LTE Saha-Boltzmann temperature sensitivity.  A
constant degree of ionization suggests dominance of constant radiation
in an optically thin environment.

\def\figlabel{fig:SBJ-cont}
\begin{figure}      
  \centering
  \includegraphics[width=88mm]{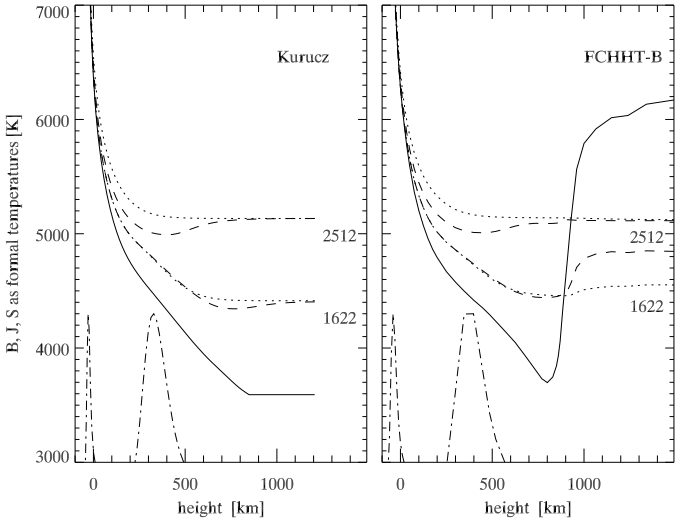} 
  \caption[]{\label{\figlabel} \footnotesize %
    The Planck function $B_\nu$ ({\em solid\/}), mean radiation
    $J_\nu$ ({\em dotted\/}), and source function $S_\nu$ ({\em
      dashed\/}) averaged over the first 
    100\,\AA\ of the \MgI\ edges
    with thresholds at $\lambda \is 2512.4$ and 1621.5\,\AA. 
    {\em Left\/}: Kurucz model.  {\em Right\/}: \fonmod\ model.
    Formal representative temperatures are plotted for these
    quantities: the electron temperature for $B_\nu$, radiation
    temperature for $J_\nu$, and excitation temperature for
    $S_\nu$. They are obtained by applying the inverse Planck function
    to $B_\nu, J_\nu, S_\nu$, respectively.  The dot-dashed
    distributions at the bottom are the intensity contribution
    functions $j_\nu\,\exp(-\tau_\nu)$, with $j_\nu$ the emissivity
    and $\tau_\nu$ the optical depth,near the thresholds wavelengths.
    The lefthand deeper-formed curve is for the 2512~\AA\ edge.
    Each curve is scaled to its maximum value. %
  }\end{figure}

\section{Explanation}                  \label{sec:explanation}
The behavior of $b_1^{\rm Zwaan}$ for \MgI\ in Fig.~\ref{fig:bground}
is well-understood, in particular since the classic PhD analysis of
the comparable iron spectrum by
\citetads{1972PhDT.........7L} 
(\cf\ \citeads{1972ApJ...176..809A}; 
review by \citeads{1988ASSL..138..185R}). 
The deep dip results from the dominance of the radiation field
$\overline{J}$ in the source function $\overline{S}$ of the \MgI\
ionization edges in the violet and ultraviolet, where the bar denotes
opacity-weighted frequency averaging over the edge and $J$ is the
angle-averaged intensity over all directions.  

Figure~\ref{fig:SBJ-cont} illustrates this cause by plotting
100\,\AA-wide averages of $S_\nu$, $B_\nu$ and $J_\nu$ for the two
major \MgI\ edges against height in the Kurucz and \fonmod\ models.
The spectral averaging smooths the many superimposed line blends.
These quantities are plotted as formal temperatures in order to
combine them in single diagrams for the two wavelengths, removing the
Planck function sensitivity to wavelength (which is exponential in the
Wien approximation valid here).  Figure~\ref{fig:SBJ-cont} shows that
across the ultraviolet $S_\nu$ tends to follow $J_\nu$ at the heights
where the radiation escapes (shown by the contribution functions at
the bottom of the graphs), with $J_\nu \!\approx\! B_\nu$ for the
2512\,\AA\ edge.

Single-transition scattering with $\overline{S} \is
(1-\varepsilon)\,\overline{J} + \varepsilon \, \overline{B}$ is an excellent
approximation for these edges which scatter strongly, with complete
redistribution over the edge profile.  The steep decrease of the
collisional destruction probability $\varepsilon$ with height makes
$\overline{S}$ uncouple from $\overline{B}$ already deep in the
photosphere, at a thermalization depth much deeper than the radiation
escape depth.
The sensitivity of the \lambdop\ operator to the steepness of the
$S_\nu(\tau)$ gradient results in $J_\nu > B_\nu$ where they uncouple.
This excess produces a superthermal edge source function in the upper
photosphere, with $\overline{S} \approx (b_c/b_i)\, \overline{B}$ in
the Wien limit for an edge between level $i$ and ionization limit $c$.
Thus, the divergence between $S_\nu$ and $B_\nu$ for the edge at
1621.5\,\AA\ in Fig.~\ref{fig:SBJ-cont} translates into the divergences
between $b_c$ (\MgII) and $b_1$ (\MgI) in Fig.~\ref{fig:bground} that
measure the source function departure from LTE for this edge.
Similarly for $b_c/b_2$ in the edge at 2512.4\,\AA\ (which is a
more important ionization channel and continuum provider).

In a plane-parallel atmosphere $J_\nu$ flattens out to a constant
outer limit unless there is significant emissivity in higher layers.
The 1D models have that in their transition regions where the steeply
increasing temperature gives larger $\varepsilon$, causing renewed
coupling to $B_\nu$.  Below the transition region, the flat $J_\nu$
may cut through whatever the model specifies as chromospheric
temperatures, so that the latter are mapped into the $b_c/b_i$ ratio
when $S_\nu$ follows $J_\nu$.  In the \fonmod\ panel of
Fig.~\ref{fig:SBJ-cont} the 2512.4\,\AA\ edge shows such
behavior.  The 1621.5\,\AA\ edge has some $S_\nu$ coupling to the
higher chromospheric temperature and a yet smaller increase in
$J_\nu$, which rises slightly above the flattened-out limit value for
the Kurucz model.

All this is beautifully illustrated across the VAL3-C spectrum in
Fig.~36 of \citetads{1981ApJS...45..635V} 
and is explained at length in
\citetads{2003rtsa.book.....R}. 

There is no significant difference between the two models in
Fig,~\ref{fig:SBJ-cont} at the heights of formation of these
bound-free \MgI\ continuum contributions.  For both edges the
formation is purely photospheric.  Backradiation from the upper
chromosphere plays no role.

The conclusion from Figs.~\ref{fig:models}--\ref{fig:SBJ-cont} is that
the \fonmod\ model differs quantitatively from VAL3-C, but not
qualitatively and not in its radiation physics.  In particular, the
continuum formation at these ultraviolet wavelengths is identical
between the two models used here since the differences in
$S_\nu$ at larger heights are not sampled by the photospheric
contribution functions.

\def\figlabel{fig:bgrids}
\begin{figure}      
  \centering
  \includegraphics[width=88mm]{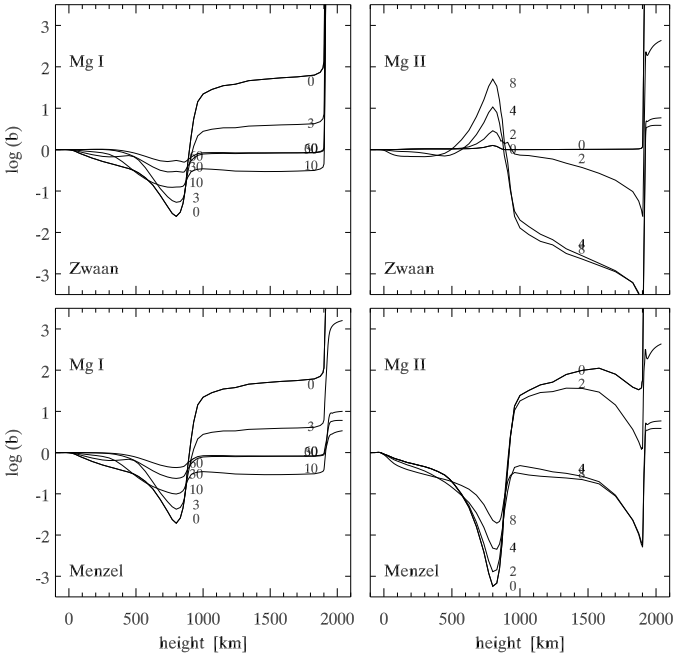} 
  \caption[]{\label{\figlabel} \footnotesize %
    Population departures of selected magnesium levels in the \fonmod\ model.
    The level numbers increase with excitation energy.
    {\em Left\/}: \MgI.  {\em Right\/}: \MgII. 
    {\em Upper row\/}: Zwaan definition. {\em Lower row\/}: Menzel definition.
  }\end{figure}

Figure~\ref{fig:bgrids} adds \fonmod\ $b$ curves for selected excited
levels in \MgI\ and \MgII\ to the ground-state display in the upper
panels of Fig.~\ref{fig:bground}.  The \MgI\ ground state has the
largest NLTE deficit in the temperature minimum due to $\overline{J} >
\overline{B}$ overionization in the principal \MgI\ edges.  The
plateau of $b_1$ (index 0) excess at larger height maps $\overline{J}
< \overline{B}$ in these scattering edges.

The higher levels in \MgI\ converge to the $b \!\approx\! 1$ value of
the \MgII\ ground state, with multi-level crosstalk
(``interlocking'').  In the \MgI\ Rydberg domain this convergence is
maintained by a collisional population replenishment flow driven by
photon losses in intermediate \MgI\ lines, exemplified by the deep
onset of the drop in the $n\is10$ curve.  Such photon-loss
replenishment from the next-ion population reservoir was called photon
suction by \citetads{1992A&A...265..237B}. 
The Rydberg flow causes the \MgI\ emission features near $\lambda \is
12$\,$\mu$m that were thought chromospheric by
\citetads{1989ApJ...340..571Z}, 
but actually are photospheric
(\citeads{1992A&A...253..567C}).  
 
The \MgII\ $b$ peaks in the second panel of Fig.~\ref{fig:bgrids}
reflect the effect of $\overline{J} > \overline{B}$ pumping in
strong ultraviolet lines, similarly as in \FeII\ (see
\citeads{1988ASSL..138..185R}). 
A particular example of the latter is the weak \FeII\,3969.4\,\AA\
line between \CaIIH\ and \Hepsilon, discovered as limb emission line
by \citetads{1929MNRAS..89..566E}. 
Its on-disk emission was initially attributed to chromospheric
backradiation by \citetads{1974A&A....33..363L}, 
but is actually caused by photospheric pumping in \FeII\ resonance
lines (\citeads{1980ApJ...241..374C}) 
which explains its extraordinary spatial intensity variation at the
limb (\citeads{1980A&AS...39..415R}). 
Similar upper-photosphere $\overline{J} > \overline{B}$
pumping sets the $b$ peak of the \MgIII\ ground state in the
first panel of Fig.~\ref{fig:bground}.

The lower panels of Figure~\ref{fig:bgrids} repeat the $b$ displays for
the Menzel definition.  The lefthand panels are the same, but on the
right the Menzel panel shows reversals of the Zwaan panel due to the
normalization by the \MgIII\ population.  It makes the \MgII\ curves
appear similar to the \MgI\ curves.

The lower panels of Figure~\ref{fig:bgrids} are similar to the
corresponding \MgI\ and \MgII\ panels in Fig.~4 of \fontenla.  The
look-alike sequences of dips, deepest for the lowest levels, in their
six Menzel plots made \fontenla\ conclude: {\em ``All our results
  display overionization (i.e., ground-level departure from LTE
  coefficients smaller than unity) of all species around the
  temperature minimum. [\ldots] Also, near the temperature minimum the
  lower levels have smaller departure coefficients than the upper
  levels, indicating that the overionization of neutrals is a result
  from FUV and/or EUV irradiation which primarily affects lower
  levels. This indicates that overionization is much more affected by
  the irradiation from the upper chromosphere in continuum and
  emission lines than by the photospheric radiation and absorption
  lines. Therefore, we stress that the consideration of a realistic
  upper chromosphere is essential to the determination of the
  densities near the temperature minimum of (1) neutral low
  first-ionization-potential (FIP) elements, and (2) singly ionized
  high FIP elements. Even a very sophisticated calculation of the
  effects of lower chromospheric absorption lines on the elemental
  ionization can produce unrealistic results if it does not include
  upper chromosphere irradiation.''}.  %
This interpretation is incorrect.  Downward irradiation does not
affect the degree of ionization of these minority species around the
temperature minimum of the \fonmod\ model.  We have demonstrated this
for Mg here; the same applies to Fe, Si, Al.

However, more positively, we wish to note that the modeling of
\fontenla\ is not impaired by this misinterpretation of its results,
and that that does not detract value from their effort.  In this era
of giant computer programs it becomes non-trivial to diagnose
what the output implies.  Chromospheric physics is becoming a key area
in solar research but does require understanding of complex
non-equilibrium optically-thick spectrum formation.  Our tutorial
elucidation of this particular issue may be instructive to newcomers
to the field.

\def\figlabel{fig:SBJ-Halpha}
\begin{figure}      
  \centering
  \includegraphics[width=88mm]{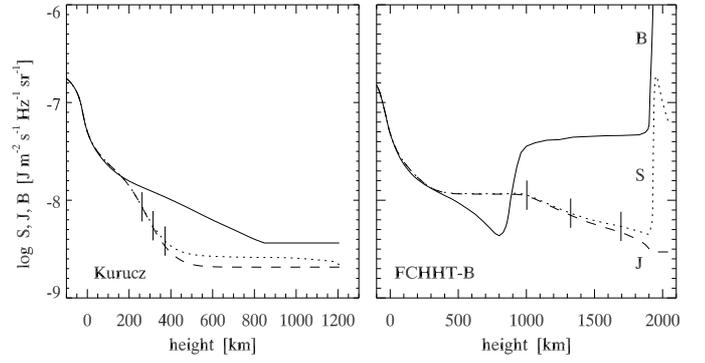} 
  \caption[]{\label{\figlabel} \footnotesize %
    \Halpha\ formation for the Kurucz ({\em left\/}) and
    \fonmod\ ({\em right\/}) models.  {\em Solid\/}: Planck function
    $B_\nu$ at line center.  {\em Dashed\/}: angle-averaged
    radiation $J_\nu$ at line center.  {\em Dotted\/}: total
    source function $S_\nu$ at line center. The ticks mark the locations
    with line-center optical depth $\tau_\nu = 3, 1, 0.3$ 
    from left to right.
  }\end{figure}

\section{Backradiation in H$\alpha$}   \label{sec:Halpha}
Figure~\ref{fig:SBJ-Halpha} shows formation parameters for \Halpha\ in
the Kurucz and \fonmod\ models.  It is similar to Fig.~8 in
\citetads{2009A&A...503..577C}, 
but here, prompted by the referee and in the vein of this
paper, we add detailed analysis using the \fonmod\ model for
demonstration.  We do not regard it a viable explanation of actual
solar \Halpha\ formation (Sect.~\ref{sec:discussion}), but use it
rather as a one-dimensional didactic ``\fonmod\ star''.

The lefthand panel of Fig.~\ref{fig:SBJ-Halpha} illustrates \Halpha\
line formation in such a star without overlying chromosphere.  At line
center, $J_\nu$ drops well below $B_\nu$.  This behavior differs from
the $J_\nu \!>\!  B_\nu$ excesses of the ultraviolet continua in
Fig.~\ref{fig:SBJ-cont} because the Planck-function sensitivity to
temperature is smaller at longer wavelengths, flattening the $\rmd
B_\nu/\rmd h$ gradient, and because the substantial additional line
opacity $\alpha^l_\nu$ flattens the $\rmd B_\nu / \rmd \tau_\nu = \rmd
B_\nu / [(\alpha_\nu^c + \alpha_\nu^l) \, \rmd z]$ gradient yet more.
The \lambdop\ operator then produces $J_\nu < S_\nu$ tending towards
the $S_\nu(0) = \sqrt{\varepsilon_\nu}B_\nu$ surface value for a
scattering isothermal atmosphere with constant $\varepsilon_\nu$.
  
In the Kurucz model the \Halpha\ core originates from the upper
photosphere around $h \is 400$\,km as a deep, strongly scattering line
with $S_\nu \approx \overline{J}_{\nu_0}$.  The corresponding emergent profile
is shown in the top panel of Fig.~\ref{fig:prof-Halpha-J}.  It is well
reproduced by applying the Eddington-Barbier approximation $I_\nu
\approx S_\nu(\tau_\nu \is 1)$.

\def\figlabel{fig:prof-Halpha-J}
\begin{figure}      
  \centering
  \includegraphics[width=75mm]{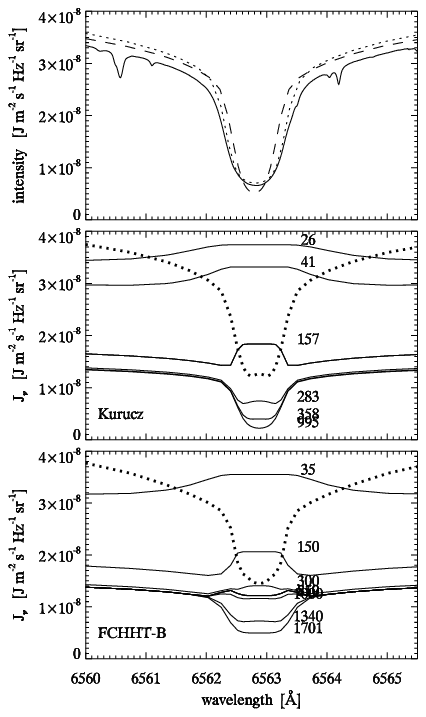} 
  \caption[]{\label{\figlabel} \footnotesize %
    {\em Top panel\/}: emergent \Halpha\ intensity profiles at disk
    center computed from the \fonmod\ ({\em dotted\/}) and Kurucz
    ({\em dashed\/}) models, in comparison with the observed
    spatially-averaged profile ({\em solid\/}) taken from the atlas
    observed by Brault \& Testerman, calibrated by
    \citetads{1984SoPh...90..205N}, 
    and posted by
    \citetads{1999SoPh..184..421N}. 
    {\em Lower panels\/}: mean intensity $J_\nu$ across \Halpha\ at
    different heights in the Kurucz and \fonmod\ atmospheres,
    respectively.  For the Kurucz panel the sampling heights are 26,
    41, 157, 283, 358 and 995~km.  For the \fonmod\ panel they are 35,
    150, 300, 519, 850, 1000, 1340 and 1701~km. The dotted profile in
    the bottom panel is the outward intensity at height 850~km,
    impinging on the \fonmod\ chromosphere. The dotted profile in the
    middle panel is the outward intensity at height 252~km in the
    Kurucz model where the line-center optical depth equals the
    optical thickness $\tau_\nu = 3.5$ of the \fonmod\ chromosphere. %
  }\end{figure}

\def\figlabel{fig:ha-opac-fon}
\begin{figure}      
  \centering
  \includegraphics[width=75mm]{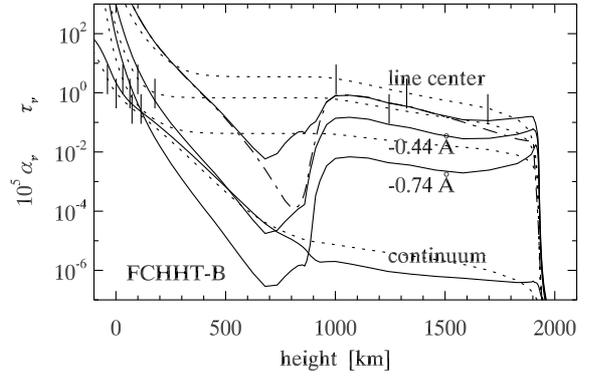} 
  \caption[]{\label{\figlabel} \footnotesize %
    {\em Solid\/}: \Halpha\ line extinction coefficients
    $\alpha_\nu^l$ per m against height for the \fonmod\ model, at
    line center and in the blue wing at $\Delta \lambda \is -0.44$ and
    $-0.74$~\AA, and the continuous opacity $\alpha_\nu^c$ in the
    adjacent continuum.  The dot-dashed curve is the line-center
    result for LTE. The opacities are multiplied by a scale height of
    $10^5$\,m to make them comparable to the corresponding optical
    depth scales ({\em dotted curves\/}).  The tick marks on the
    latter correspond to $\tau_\nu = 3, 1, 0.3$.  Across the opacity
    dip due to the temperature minimum the optical depth is $\tau_\nu
    = 3.5, 0.7, 0.04$ at the three line wavelengths, respectively.
  }\end{figure}

The righthand panel of Fig.~\ref{fig:SBJ-Halpha} shows the formation
of \Halpha\ in the \fonmod\ model.  Its chromospheric high-temperature
plateau supplies sufficient \Halpha\ opacity that at line center
$\tau_\nu \is 1$ is reached about a thousand km higher than in the
Kurucz model. The line-center formation is again well described by
$I_\nu \approx S_\nu(\tau_\nu \is 1) \approx
\overline{J_{\nu_0}}(\tau_\nu \is 1)$, but the Eddington-Barbier
approximation is less well applicable in the line wings which have a
formation gap between photosphere and chromosphere (\eg\
\citeads{1972SoPh...22..344S}; 
\citeads{2006A&A...449.1209L}). 
This gap is detailed in Fig.~\ref{fig:ha-opac-fon} which shows
\Halpha\ opacities and optical depth scales, with the opacities
multiplied by the scale height in the upper \fonmod\ photosphere (it
doubles in the \fonmod\ chromosphere from a steep increase of the
imposed non-gravitational acceleration). The line extinction has a
deep dip in the temperature minimum.  Correspondingly, the optical
depth buildup (dotted curves) levels out.  At line center the
\fonmod\ chromosphere has optical thickness $\tau_\nu \is
3.5$. At $\Delta \lambda \is -0.44$~\AA\ $\tau_\nu \is 0.3$ (outer
tick) is reached already in the chromosphere but $\tau_\nu \is 1$ only
a thousand km deeper in the photosphere.  At $\Delta \lambda =
-0.74$~\AA\ the \fonmod\ chromosphere is virtually transparent.

At line center the line extinction much exceeds the continuous
extinction even in the dip, so that the total source function equals
the line source function at all heights.  In the wings the line
extinction drops below the continuum extinction.  Where this happens
the total source function drops to the Planck function (not shown).

\def\figlabel{fig:ha-betas-fon}
\begin{figure}      
  \centering
  \includegraphics[width=75mm]{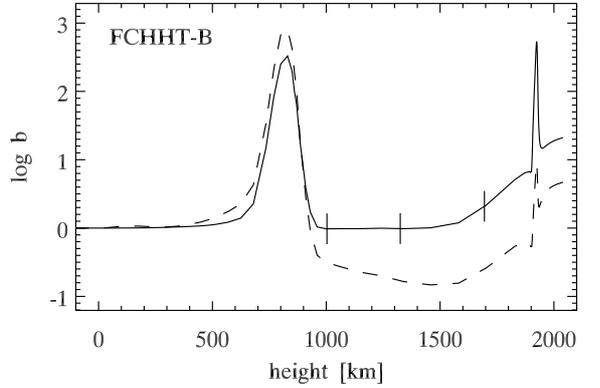} 
  \caption[]{\label{\figlabel} \footnotesize %
    \Halpha\ population departure coefficients against height for the
    \fonmod\ model. The tick marks correspond to $\tau_\nu = 3, 1,
    0.3$.  {\em Solid\/}: lower-level departure coefficient $b_2$.
    {\em Dashed\/}: upper-level departure coefficient $b_3$. %
  }\end{figure}
 
Figure~\ref{fig:ha-betas-fon} shows the departure coefficients $b_2$
and $b_3$ (Zwaan definition) governing \Halpha\ in the \fonmod\ model.
The curve divergence $\log(b_3)-\log(b_2)$ corresponds to the curve
divergence $\log(S_\nu) - \log(B_\nu)$ in the \fonmod\ panel of
Fig.~\ref{fig:SBJ-Halpha}.  Throughout the near-isothermal \fonmod\
chromosphere the \Halpha\ opacity is nearly in LTE because the
formation of \Lyalpha\ is close to detailed balancing there.  However,
this is not the case in the deep temperature minimum and steep rise to
the chromospheric temperature.  Here the \lambdop\ operator smooths
the rapid temperature changes for the scattering-dominated \Lyalpha\
source function (not shown) which has $\overline{S_\nu} \approx
\overline{J_\nu} \approx b_2\, B_\nu$.
This \Lyalpha\ smoothing causes the high peak in $b_2$ and
considerable corresponding fill-in of the opacity dips in
Fig.~\ref{fig:ha-opac-fon}, illustrated for line center as difference
with the much deeper dot-dashed LTE curve.  In this manner \Lyalpha\
scattering compensates for low-temperature \Halpha\ opacity loss when
cool features are narrow in spatial
extent.  The $b_2$ peak in Fig.~\ref{fig:ha-betas-fon} doubles if CRD
is adopted for \Lyalpha, from additional smoothing through farther
wing-photon travel.

The corresponding emergent intensity profile in the top panel of
Fig.~\ref{fig:prof-Halpha-J} is similar to the Kurucz result despite
the disparate line formation. Both are reasonably good approximations
to the observed mean quiet-Sun profile.  Both gain fit quality in the
wings if the computed continuum intensity is rescaled to the observed
value. Increase of the microturbulence in the Kurucz model from 1.5 to
about 10~\kms\ produces a good core fit for this model as well.

The $J_\nu$ curves in Fig.~\ref{fig:SBJ-Halpha} show that the opacity
gap in the \fonmod\ model contains much more \Halpha\ radiation than
at corresponding heights in the Kurucz model.  This \Halpha\ radiation
fill-in causes a yet higher peak for $b_3$ in
Fig.~\ref{fig:ha-betas-fon}.  It is detailed in the lower panels of
Fig.~\ref{fig:prof-Halpha-J}. In the lower photosphere the \Halpha\
$J_\nu$ profiles are the same for the two models, but above $h \approx
250$~km the Kurucz model gives absorption cores whereas the \fonmod\
model predicts slightly self-reversed cores across the opacity gap.
This difference explains that for the Kurucz model
$\overline{J}_{\nu_0}$ (shown by the line-center $S_\nu$ curve in the
lefthand panel of Fig.~\ref{fig:SBJ-Halpha}) ends up higher than
$J_\nu$ at line center, whereas for the \fonmod\ model $S_\nu \approx
\overline{J}_{\nu_0}$ coincides with line-center $J_\nu$ across the
opacity gap in the righthand panel of Fig.~\ref{fig:SBJ-Halpha}.  The
large core width of the highest-formed \fonmod\ profiles comes from
large thermal broadening (about 10~\kms) and yet larger
microturbulence (about 15~\kms).

Because the \fonmod\ chromosphere is near-isothermal the \Halpha\
formation is comparable to the classic results for a finite isothermal
scattering atmosphere of
\citetads{1965MNRAS.130..295A}, 
in particular their optically thick but effectively thin case in which
the line source function does not reach thermalization.  The main
difference is that the \fonmod\ chromosphere is irradiated from below.
The impinging intensity profile in the outward direction is shown by
the dotted curve in the bottom panel of Fig.~\ref{fig:prof-Halpha-J}.
This line is much shallower than the emergent profile from the Kurucz
model, in agreement with the much higher $J_\nu$ above $h \!\approx\!
200$~km.  A test with a hotter chromosphere, \ie\ thicker in \Halpha,
gives more backradiation and a higher impinging profile with a bright
self-reversed core.  However, the deep emergent-intensity core remains
about the same.

Thus, the presence of an opaque chromosphere changes the
illumination profile with respect to the profile that emerges from the
same photosphere without overlying chromosphere.  A mathematical
explanation is given by the principle of invariance (page 165 of
\citeads{1950ratr.book.....C}) 
regarding addition or subtraction of a layer of arbitrary optical
thickness to a semi-infinite plane-parallel atmosphere; creating a gap
or pushing down the finite chromospheric atmosphere over the gap to
where it meets similar source conditions does not change the emergent
radiation.  A more physical explanation is that the radiation field in
the gap builds up due to backradiation from the overlying chromosphere
(called reflection by \citeads{1950ratr.book.....C}). 

\def\figlabel{fig:S2level-fon}
\begin{figure}      
  \centering
  \includegraphics[width=75mm]{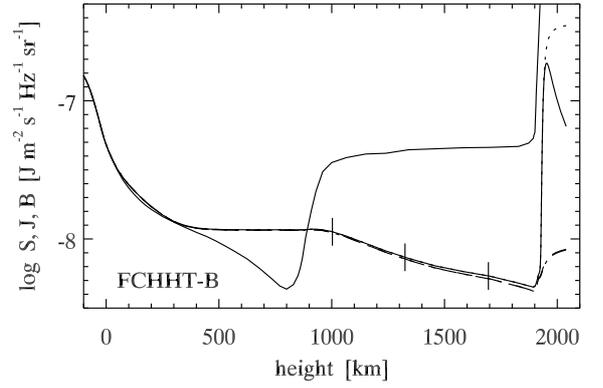}  
  \caption[]{\label{\figlabel} \footnotesize %
    \Halpha\ source function comparisons for the \fonmod\ model.  {\em
      Solid\/}: Planck function and total source function at line
    center.  The latter has ticks at $\tau_{\nu_0} = 3, 1, 0.3$.  {\em
      Dotted\/}: line source function $S^l_{\nu_0}$, differing from
    the total source function only at the very top of the atmosphere.
    {\em Dashed and dot-dashed\/}: two-level approximation 
    $(1 - \varepsilon_{\nu_0}) \, \overline{J}_{\nu_0} +
    \varepsilon_{\nu_0} B_{\nu_0}(T)$ and profile-averaged mean
    intensity $\overline{J}_{\nu_0}$.  These coincide everywhere.
  }\end{figure}

We now demonstrate that this backradiation is dominated by two-level
scattering.  The line source function for complete redistribution can
be written as:
\begin{eqnarray}
  S^l_{\nu_0} &=& (1 - \varepsilon_{\nu_0} - \eta_{\nu_0}) \,  
  \overline{J}_{\nu_0}
  + \varepsilon_{\nu_0}  B_{\nu_0}(T) + \eta_{\nu_0}  B_{\nu_0}(T_\rmd) \\
  &=& \frac{\overline{J}_{\nu_0} 
    + \varepsilon_{\nu_0}^\prime B_{\nu_0}(T)
    + \eta_{\nu_0}^\prime B_{\nu_0}(T_\rmd)}
  {1 + \varepsilon_{\nu_0}^\prime + \eta_{\nu_0}^\prime}, 
  \label{eq:Sepsprime}
\end{eqnarray}
where $\overline{J}_{\nu_0}$ is the profile-averaged mean intensity,
$\varepsilon_{\nu_0}$ the thermal destruction probability, \ie\ the
fraction of line-photon extinctions by direct collisional
deexcitation corrected for emissivity from spontaneous
scattering, $\eta_{\nu_0}$ the fractional probability of all
indirect detour extinction, with detour meaning any multi-level path
from the upper to the lower level not including the direct transition,
corrected for stimulated detour emissivity,
and $\varepsilon_{\nu_0}^\prime$ and $\eta_{\nu_0}^\prime$ the
corresponding ratios of such extinctions to the contribution by
two-level scattering.  The formal detour excitation temperature
$T_\rmd$ is given by $(g_u \, D_{ul})/(g_l \, D_{lu}) \equiv
\exp(h\nu_0/kT_\rmd)$ where $D_{ul}$ is the summed transition
probability (per second per particle in the upper level) of all detour
upper-to-lower paths, $D_{lu}$ for all detour lower-to-upper paths,
and $g_l$ and $g_u$ are the statistical weights.

The classic literature used Eq.~\ref{eq:Sepsprime} (\eg\
\citeads{1974ApJ...188..399G}; 
Sect.~8.1 of \citeads{1968slf..book.....J}; 
Eq.~12.11 of
\citeads{1970stat.book.....M}) 
following \citetads{1957ApJ...125..260T} 
who divided strong lines into ``collision type'' with
$\varepsilon_{\nu_0} > \eta_{\nu_0}$ and ``photoelectric type'' with
$\eta_{\nu_0} > \varepsilon_{\nu_0}$, designating \Halpha\ as
principal example of the second type.  In particular, \Halpha\ would
gain most photons from Balmer ionization followed by a recombination
path into $n \is 3$ employing photospheric Balmer photons (\eg\
\citeads{1961psc..book.....A}; 
see also \citeads{1985cdm..proc..288C}). 
The corresponding schematic source function diagram in Fig.~3 of
\citetads{1959ApJ...129..401J} 
(reprinted in Fig.~12-9 of
\citeads{1970stat.book.....M} 
and Fig.~11-11 of
\citeads{1978stat.book.....M}) 
is qualitatively similar to the \fonmod\ \Halpha\ behavior in
Fig.~\ref{fig:SBJ-Halpha}, with $S_\nu^l$ leveling out in the
photosphere to become much higher than $B_\nu$ in the temperature
minimum.

\def\figlabel{fig:eps-eta-fon}
\begin{figure}      
  \centering
  \includegraphics[width=75mm]{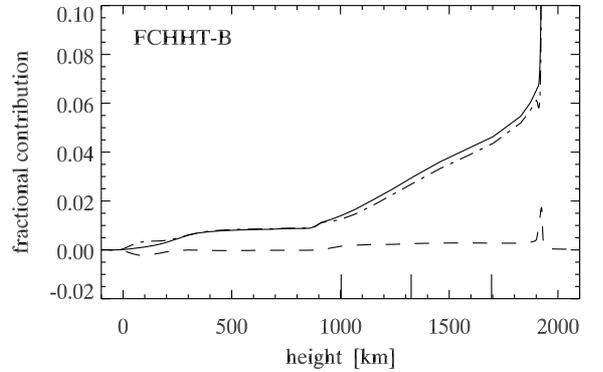} 
  \caption[]{\label{\figlabel} \footnotesize %
    Fractional collisional contribution $\varepsilon_{\nu_0}
    [B_{\nu_0}(T) \!-\!  \overline{J}_{\nu_0}]/S^l_{\nu_0}$ ({\em
      dashed\/}) and detour contribution $\eta_{\nu_0}
    [B_{\nu_0}(T_\rmd) \!-\!  \overline{J}_{\nu_0}]/S^l_{\nu_0}$ ({\em
      dot-dashed\/}) to the \Halpha\ line source function.  Their sum
    ({\em solid\/}) represents
    $(S^l_{\nu_0}-\overline{J}_{\nu_0})/S^l_{\nu_0}$.  The ticks on
    the $x$-axis are at $\tau_{\nu_0} \is 3, 1, 0.3$ in \Halpha.
  }\end{figure}

However, in the \fonmod\ case this behavior is due to the 
backscattering from the chromosphere, not from Balmer-continuum detour
emission.  If the latter were important it would raise the \Halpha\
source function also for the Kurucz model.  The Balmer continuum
scatters outward similarly to the 2512~\AA\ continuum in
Fig.~\ref{fig:SBJ-cont}, the same in both models up to $h \approx
800$~km, with about 1400~K radiation temperature excess in the 
temperature minimum -- but that region is transparent in \Halpha.

The scattering nature of the \Halpha\ source function is demonstrated
in Fig.~\ref{fig:S2level-fon} by comparing it with the mean radiation
$\overline{J}_{\nu_0}$ and the two-level approximation.  They are
closely the same everywhere.  The differences are magnified in
Fig.~\ref{fig:eps-eta-fon} showing the fractional collision and detour
contributions.  The collisional contribution is negligible except in
the deep photosphere where collisional excitations create most
\Halpha\ photons (with $\varepsilon_{\nu_0} \!\approx\! 1$ and
$\overline{J}_{\nu_0} \!\approx\! B_{\nu_0}$).  The latter
resonance-scatter outward.  Higher up, the detour contribution is much
larger, confirming the $\eta_{\nu_0}/\varepsilon_{\nu_0}$ evaluation
of \citetads{2004A&A...418.1131A}, 
but it still amounts to only a few percent of the line source function
and raises the line-center intensity by only one percent of the
continuum intensity.  In the steep temperature rise of the \fonmod\
transition region the detour contribution grows rapidly to nearly
100\%, but at negligible \Halpha\ opacity and therefore of no
importance to the emerging profile -- and not due to photospheric
irradiation.

The conclusion from this section is that in the \fonmod \ model
\Halpha\ is first and foremost a scattering line, with chromospheric
backscattering boosting the line-center radiation across the opacity
gap between photosphere and chromosphere. The \Halpha\ core emerges
from the \fonmod\ chromosphere but most photons were created in the
deep photosphere.  For \Halpha\ the \fonmod\ chromosphere is primarily
a scattering attenuator that builds up its own irradiation from below.
The main source function difference with other strong scattering lines
such as the \CaII\ resonance lines and infrared triplet is the
effect of the opacity gap, not the amount of detour contribution.

\section{Discussion}                   \label{sec:discussion}
Static 1D modeling of the continua from the solar atmosphere has
reached such sophistication that even the modelers themselves may
misinterpret their results (Sect.~\ref{sec:explanation}).
Interpretation of chromospheric fine structure using strong lines, in
particular \Halpha, has relied more on cloud modeling (see the
excellent review by \citeads{2007ASPC..368..217T}) 
than on static model-atmosphere interpretation as in
Sect.~\ref{sec:Halpha}.  In the meantime, observations of the actual
chromosphere (Sect.~\ref{sec:observation}) have reached such spatial
and temporal resolution that the worlds of the modelers and the
observers now seem far apart.  A third worldview is given by
time-dependent numerical MHD simulations of the chromosphere (\eg\
\citeads{2006ASPC..354..345S}; 
\citeauthor{2007A&A...473..625L} 
\citeyearads{2007A&A...473..625L}, 
\citeyearads{2010ApJ...709.1362L}; 
\citeads{2011ApJ...736....9M}). 
Obviously, the different views should come together.  It falls
outside the scope of this paper to effectuate such synthesis, but we
make a few salient points.

\paragraph{Emergent spectrum fitting.}
The \fonmod\ model and the AL-C7 model of
\citetads{2008ApJS..175..229A} 
fit the ultraviolet atlases about equally well, but differ appreciably
in their stratifications (Fig.~\ref{fig:models}) and in their
stratification physics.  Both assume hydrostatic equilibrium, but the
AL-C7 model gains additional chromospheric support from imposed
turbulent pressure inspired by observed line broadening, as in its VAL
and FAL predecessors.  The \fonmod\ model gains chromospheric support
instead from imposed non-gravitational acceleration inspired by the
Farley-Buneman instability
(\citeads{2008A&A...480..839F}). 
Thus, best-fit model determination from such data does not represent a
unique data inversion.  

An extreme example of non-uniqueness is that
the two disparate \Halpha\ formations in Fig.~\ref{fig:SBJ-Halpha},
with and without a chromosphere, can both fit the observed average
\Halpha\ profile (top panel of Fig.~\ref{fig:prof-Halpha-J}).
Figure~\ref{fig:DOT-SDO} obviously suggests that chromospheric
\Halpha\ formation as in the \fonmod\ model is more realistic than
photospheric \Halpha\ formation as in the Kurucz model, but the
predicted \Halpha\ profile cannot be used as discriminator.

\paragraph{One-dimensional continuum modeling.}
The similarity of the magnetogram, 1700\,\AA\ and \CaIIH\ panels in
Fig.~\ref{fig:DOT-SDO} suggests that 1D modeling along vertical
columns which differs between internetwork, network, and plage may be
a reasonable approximation for ultraviolet continua from these
features.  The bright grains that constitute network and plage mark
strong-field structures that are roughly vertical.  However, facular
modeling towards the limb must then account for slanted viewing across
such structures by cutting through different 1D models along the line
of sight, as in the classic Z\"urich fluxtube modeling of \eg\
\citetads{1993A&A...268..736B}. 

\paragraph{Cloud modeling.}
The \fonmod\ chromosphere may be regarded as an horizontally extended
isothermal slab but not as a constant-property cloud since it is more
like a finite atmosphere, with density stratification, outward $S_\nu$
decay, and Eddington-Barbier-like \Halpha\ core formation.  The source
function is dominated by scattering (Fig.~\ref{fig:S2level-fon}) and
the slab is effectively thin (Fig.~\ref{fig:ha-opac-fon}) so that the
source function amplitude depends on the slab opacity. Due to
backscattering, the \Halpha\ profile illuminating the slab from below
(dotted curve in the bottom panel of Fig.~\ref{fig:prof-Halpha-J}) is
shallower than for locations without an overlying chromospheric
blanket (Kurucz prediction in the top panel of
Fig.~\ref{fig:prof-Halpha-J}; \cf\
\citeads{2010MmSAI..81..769B}) 
and brightens for a thicker blanket.  If actual fibrilar canopies
share such properties than these upset classical cloud modeling.  

At the suggestion of the referee, we added in the middle panel of
Fig.~\ref{fig:prof-Halpha-J} the outward intensity profile at that
optical depth in the Kurucz model which equals the optical thickness
of the \fonmod\ chromosphere.  It is similar to the outward intensity
profile impinging the latter (bottom panel) because the line formation
is much more similar for the two models on optical depth scales than
on geometrical height scales.  The near-equality of these profiles
suggests a new recipe for 1D \Halpha\ cloud modeling, namely to use
the outward intensity profile in a radiative-equilibrium photosphere
at optical depth equal to the cloud's thickness as impinging
background profile.

\paragraph{Dynamics.}
The actual internetwork atmosphere above the photosphere (in our view
consisting of a cool clapotisphere, chromospheric fibril canopy, and a
similar-morphology sheath-like transition region) is continuously
shocked.  In \Halpha\ ubiquitous internetwork shocks are evident in
Doppler timeslices (\eg\
\citeads{2008SoPh..251..533R}; 
\citeads{2009A&A...503..577C}). 
In magnetic concentrations shocks are ubiquitous already in the
photosphere.  Shocks also produce the dynamic fibrils jutting out from
network (\citeads{2006ApJ...647L..73H}; 
\citeads{2007ApJ...655..624D}). 
The elusive straws\,/\,spicules-II\,/\,RBEs near network are even more
dynamic.  This basic chromospheric dynamism upsets static chromosphere
modeling.

\paragraph{Departures from LTE.}
The \fonmod\ NLTE departures discussed here reach two to three orders
of magnitude at the \fonmod\ temperature minimum
(Figs.~\ref{fig:bground}, \ref{fig:ha-betas-fon}) while the lower
level of \Halpha\ is nearly in LTE where the line forms
(Fig.~\ref{fig:ha-betas-fon}).  The non-equilibrium simulation of
\citetads{2007A&A...473..625L} 
predicts overpopulations of this level up to {\em twelve\/} orders of
magnitude for cool post-shock phases, where the temperature may fall
well below the temperature minima of the standard models
(\citeads{2011A&A...530A.124L}). 
Thus, dynamic behavior may dramatically increase the temperature
variations and NLTE effects.

\paragraph{Height of the transition region.}
One-dimensional models for different features typically differ
primarily in the height of their transition region.  Simulations as
the one of \citetads{2007A&A...473..625L} 
suggest that the transition region above internetwork is kicked up and
mass-loaded by field-guided flows that do not obey one-dimensional
hydrostatic equilibrium.  In this simulation the location of the
transition region varies over $h\is1000-4000$\,km, with rapid changes,
and is generally lower above magnetic concentrations.  Such larger
range than in the 1D model grids may better reflect actual variations
in fibril-canopy height or corrugation.

\paragraph{ \Halpha\ opacity gap.}
The \Halpha\ opacity gap and backscattering into it are properties of
the \fonmod\ model.  Do they also occur in the real Sun?  The images
in Fig.~\ref{fig:DOT-SDO} roughly agree with \fonmod\ predictions: in
the 1700~\AA\ image no fibrils are seen whereas they appear opaque in
the \Halpha\ image.  Also, when one samples the real Sun in \Halpha\
away from line center (\eg\ in DOT
movies\footnote{\url{http://www.staff.science.uu.nl/~rutte101/dot}})
the photospheric granulation appears when the fibrils become
transparent, without an intermediate clapotispheric scene as in the
\CaIIH\ and 1700\,\AA\ images in Fig.~\ref{fig:DOT-SDO}.  Thus, a
similar opacity gap seems to exist under the actual fibrilar canopies
making up the internetwork chromosphere.  The radiation crossing it,
boosted by backscattering, will suffer substantial spatial smoothing
of the photospheric scene.  Some of it may leak out through an
effectively thin fibril canopy, scatter around dark fibrils in photon
channeling as suggested by
\citetads{2004A&A...418.1131A}, 
or be seen from aside as in the backradiation explanation of bright
rims under filaments (\eg\
\citeads{1975SoPh...45..119K}; 
\cf\ \citeads{2010MmSAI..81..673P}). 
 
\paragraph{\Halpha\ detour brightening.}
In the \fonmod\ model \Halpha\ is an almost pure scattering line
(Fig.~\ref{fig:S2level-fon}). The detour contribution $\eta_{\nu_0}
(B_{\nu_0}(T_\rmd) - \overline{J}_{\nu_0})$ to the line source
function becomes important only in the transparent transition region
(Fig.~\ref{fig:eps-eta-fon}).  In the real Sun locations with very
bright \Halpha, such as the moss and active-region heart in Fig.~13 of
\citetads{2007ASPC..368...27R}, 
may represent low-lying, denser transition regions that radiate
\Halpha\ through recombination paths.


\section{Conclusion}                   \label{sec:conclusion}
The development of standard models of the solar atmosphere,
masterminded by E.H.~Avrett, represents a well-established pinnacle of
sophistication with respect to the application of NLTE spectrum
formation theory with the inclusion of numerous spectral features.
However, the assumption of hydrostatic and time-independent equilibria
without magnetism remains a far cry from the actual solar
chromosphere, which is pervaded by shocks and rapidly changing
magnetic fine structure in even the quietest regions.

Conversely, state-of-the-art magnetohydrodynamics simulations do a
good job in emulating the small-scale magnetodynamism of the actual
solar atmosphere, but they remain weak in properly treating
non-equilibrium radiation.  Implementation of the art of the 1D
modelers into the 3D time-dependent codes of the simulators presents a
formidable challenge, but seems the most promising venue to understand
the enigmatic solar chromosphere (\cf\ 
\citeads{2012arXiv1202.1926L}). 

\begin{acknowledgements}
  We thank E.~Romashets, A.~Sukhorukov and P.~S\"utterlin and the SDO
  team for their contributions to Fig.~\ref{fig:DOT-SDO}, H.~Wang for
  discussing bright filament rims, and J.~Fontenla and the referee for
  pointing out severe shortcomings in earlier versions.  This work was
  started at the Lockheed-Martin Solar and Astrophysics Laboratory
  when both authors were visiting, supported by NASA contracts
  NNG09FA40C (IRIS) and NNM07AA01C (HINODE).  Our research made much
  use of NASA's Astrophysics Data System.
\end{acknowledgements}




\end{document}